\documentclass[twocolumn,preprintnumbers,amsmath,amssymb,showpacs]{revtex4}

\pdfoutput=1 

\usepackage{graphicx}
\usepackage{bm}
\usepackage{dcolumn}
\usepackage{amssymb}
\usepackage{amsmath}
\usepackage{amsfonts}
\usepackage{color}
\usepackage{epsfig}

\def \half{{1 \over 2}}

\def \Z2{$\mathbb{Z}_2$}
\def \Y1{$\mathbb{Z}$}
\def \da{^{\dagger}}
\def \PNe{P_{N,\epsilon}}
\def \Pe{P_{\epsilon}}
\def \A{\mathcal{A}}
\def \B{\mathcal{B}}
\def \erf{\textrm{erf}}

\begin{document}
\draft

\title{
    Determining topological order from a local ground-state correlation function }
\author{
    Zohar Ringel and Yaacov E.~Kraus }
\affiliation{
     Department of Condensed Matter Physics, Weizmann Institute of Science, Rehovot 76100, Israel}

\begin{abstract}
Topological insulators are physically distinguishable from normal insulators only near edges and defects, while in the bulk there is no clear signature to their topological order. In this work we show that the $\mathbb{Z}_2$ index of topological insulators and the $\mathbb{Z}$ index of the integer quantum Hall effect manifest themselves locally. We do so by providing an algorithm for determining these indices from a local equal time ground-state correlation function at any convenient boundary conditions. Our procedure is unaffected by the presence of disorder and can be naturally generalized to include weak interactions. The locality of these topological indices implies bulk-edge correspondence theorem.
\end{abstract}

\maketitle

\section{Introduction} 
\label{Sec:Intro}

The theoretical proposal and experimental discovery of topological band insulators \cite{review} (TI) has been raising increasing interest in the condensed matter physics community. These materials form a novel topological state of matter, which does not fall into the standard classification of broken symmetries. Instead, what distinguishes this phase from a trivial band insulator (BI) is the existence of a nontrivial \Z2 topological index associated with the full band structure. This distinction is somewhat analogous to the integer quantum Hall effect (IQHE), whose distinguished states can be attributed to a \Y1 topological index associated with full Landau levels of noninteracting electrons.

Broken symmetry phases are characterized by local order parameters, in what is known as the Landau paradigm \cite{LL}. It is commonly accepted that this paradigm does not apply to the above topological phases. Instead, such phases are described by the more elusive quantity, known as a topological order \cite{definition}.

In IQHE the quantized Hall conductance is a direct manifestation of the topological order, and for the case of free electrons, this response function is a local bulk quantity \cite{NIULOCAL}. As far as we know, for TI the implication of the topological order is only through edge effects (see, for example, Refs.~\cite{SHOU-SCIENCE,KERR,HASAN}) or defect-related effects \cite{ASHVIN1}. There is no local response function that is known to characterizes this \Z2 phase, let alone an order parameter.

In this work we show that both the \Z2 index of the TI and the \Y1 index of the IQHE, as well as any gapped insulator with non-zero Chern number $\nu$, can be extracted from a local equal time ground-state correlation function. This implies that \emph{in these systems the topological order is in fact a local ground state property}. It also proves a bulk-edge correspondence theorem for such local topological orders.

When interactions are taken into account, the \Y1 index remains well defined \cite{NIU}, while it is yet unclear whether the \Z2 index does \cite{KITAEV,VOLOVIK}. Our formulation, however, remains well defined at least for weak interactions, and thus suggests an extension to the definition of the TI to weakly interacting systems. Beyond weak interactions, we show that either the energy gap or some ``occupation gap" must be closed during the transition from a TI to a BI.

The theoretical procedure that we provide can be straightforwardly adapted to form an algorithm for numerically determining the \Z2 and \Y1 indices. This algorithm is rather efficient, since one only needs to diagonalize several matrices with dimensions of the order of the correlation length squared (2D) or cubed (3D). Such an algorithm may help in numerically testing a predicted topological order.

\section{Finding locally the topological order} 
\label{Sec:Local}

Our procedure of extracting the topological index deals with noninteracting
lattice-based band insulators. The required input is the equal time ground-state
two-point correlation function
\begin{align} \label{Eq:PofGS}
P_{ij} = \langle gs | \psi\da_i \psi_j | gs \rangle,
\end{align}
where $|gs\rangle$ is the many body ground state, and $\psi\da_i$ ($\psi_i$) is the
creation (annihilation) operator of an electron in site $i$ (for brevity we let $i$
encompass also the orbital and spin degrees of freedom). For noninteracting
electrons the two-point correlation function coincides with the single-particle
spectral projector
\begin{align} \label{Eq:Pij}
P_{ij} = \sum_{E_n < \mu} \langle i | n \rangle \langle n | j \rangle,
\end{align}
where $|i\rangle$ is the single-particle wave function associated with $\psi\da_i$,
$|n\rangle$ is an eigenstate of the single-particle Hamiltonian with an energy
$E_n$, and $\mu$ is the chemical potential. Since we are interested in local bulk
properties, we assume that the sites $i$ and $j$ reside within the bulk.

In the following (see Property I below) we show that $P_{ij}$ is exponentially
local, namely it decays exponentially with the distance between $i$ and $j$ within
some correlation length $l_p$ and has only an exponentially small dependence on
details of the Hamiltonian outside a local region around $i$ and $j$. This locality
allows us to discuss $P_{ij}$ within some given $\A$, without concerning
ourselves with details of the edges of the region, boundary conditions, or the
Hamiltonian outside $\A$. In particular, for any geometry of $\A$, we can always
consider a subregion with a geometry of a Corbino disk, or a thick torus (Corbino donut) in 3D,
which we assume to have circumferences larger than $l_p$.

Given $P$ on such a region, we multiply $P_{ij}$ by a geometric phase factor,
in a way that we call ``artificial flux insertion''
\begin{align} \label{Eq:PofPhi}
P_{ij}(\phi) = P_{ij} e^{ i\theta_{ij}\phi },
\end{align}
where $\phi \in \mathbb{R}$ and $\theta_{ij} \in (-\pi, \pi]$ is the azimuthal angle from site
$i$ to site $j$. In the proceeding we assume for convenience that in 2D $\A$ is of the
topologically equivalent cylindrical geometry.

A key analytical result in this work (Property II) is that Eq.~(\ref{Eq:PofPhi}) captures
the effect of real magnetic flux insertion, up to $O(e^{-L/l_p})$ corrections, where $L$
is the inner circumference of the Corbino disk (or donut). Hence the inclusion of flux
is merely a transformation of $P$, which yields no extra information in addition to
the information already contained in $P$ \cite{Chern}.

The next step is to construct 1D Wannier functions out of $P$.
Let $X$ be the position operator along the open coordinate $\hat{x}$. We define the 1D Wannier
functions $|w_n (\phi) \rangle$ to be the eigenstates of the projected position
operator in a given flux \cite{KIVELSON}
\begin{align} \label{Eq:PXP}
& PXP|_\phi |w_n(\phi)\rangle = x_n(\phi) |w_n(\phi)\rangle.
\end{align}

This definition of the Wannier functions has several advantages. First, it relies on
$P$ rather than the Bloch wave functions, hence the eigenvalues ${x_n}$ are
unaffected by the phase freedom in Fourier space, on the one hand, and it remains
defined in the presence of disorder, on the other hand. Second, since $P(\phi)$ is a
continuous function of $\phi$, the $x_n$'s are also continuous in $\phi$. Third, in
time-reversal-preserving systems $PXP|_{\phi}$ is a time-reversal-invariant operator
for $\phi = 0, \pi$, and thus Kramers' theorem assures that the Wannier functions
come in time-reversal pairs with doubly degenerate eigenvalues at these points.
Last, we prove below (Property III) that within the bulk each Wannier function is
exponentially localized around its eigenvalue along $\hat{x}$, even for cases of
nontrivial Chern number. This implies that the $x_n$'s of the Wannier functions
that reside within the middle of the region $\A$ are unaffected by details
of the system out of $\A$ or by the edges of $\A$.

So far we have shown that given $P(\phi=0)$ in a local region $\A$, one has
sufficient data to extract ${x_n(\phi)}$. In order to extract the \Z2 index out of
the $x_n(\phi)$'s, we follow Ref.~\cite{FUKANE} and consider the pairs of
eigenvalues at $\phi = 0$ and $\phi = \pi$. As mentioned before, at these fluxes the
$x_n$'s come in degenerate pairs. The difference between a BI and a TI is whether
the pairs at $\phi = 0$ are the same as those at $\phi = \pi$ (BI) or not (TI), as
depicted in Fig.~\ref{Fig:Switching}.

The \Y1 index is even simpler to extract, since there is no degeneracy in the $x_n$'s, and all the 1D Wannier functions move in the same direction \cite{ZAK}. According to gauge invariance, the eigenvalues at $\phi = 0$ are the same as those at $\phi = 2\pi$, but each $x_n$ may be carried with the flux to $x_m$. If the labeling of the eigenvalues is such that $x_{n+1} > x_n$, then the \Y1 index equals to $m - n$, as depicted in Fig.~\ref{Fig:Z_index}.

The four \Z2 indices of the 3D TI $(\nu_0; \nu_x, \nu_y, \nu_z)$ can be extracted by generalizing the pair switching criterion \cite{FUKANEMELE}. Given a sample with periodic boundary conditions in $\hat{y}$ and $\hat{z}$, the Wannier centers are carried with two independent fluxes $x_n(\phi_y, \phi_z)$, where $\phi_y$ ($\phi_z$) corresponds to the phase twist in $\hat{y}$ ($\hat{z}$). Now $\nu_y = 1$ if the pairs switch partners between $(\phi_y, \phi_z) = (\pi,0)$ and $(\pi,\pi)$ and, accordingly, $\nu_z = 1$ for switching between $(0,\pi)$ and $(\pi,\pi)$. If the pair switching in the course $(0,0) \rightarrow (0,\pi)$ differs from that in $(\pi,0) \rightarrow (\pi,\pi)$, then $\nu_0 = 1$. In order to extract $\nu_x$ we must have $P$ also in a geometry that is periodic in $\hat{x}$ and $\hat{y}$.

We can see that in 3D TI the geometry on which $P$ is given plays an important role. The 3D generalization of the 2D cylinder is periodic boundary conditions in two directions, while the 3D generalization of the Corbino disk is the Corbino donut. The Corbino donut can be isolated from any given sample by discarding sites from the projector, and therefore properties inferred from these geometries must be local and isotropic. On the contrary, the 3D cylinder cannot be isolated form larger samples, and therefore properties inferred from this geometry need not be local nor isotropic. In the Corbino donut we consider only the flux that resides inside the donut, and $X$ denotes the open radial coordinate. For $\nu_0 = 1$ we expect pair switching, since gapless states will appear at the surface \cite{FUKANEMELE}. On the other hand, since a Corbino donut can be isolated from a larger sample in any orientation, it is impossible to extract from it the orientation-dependent weak indices.

In the presence of interactions, the ground-state two-point correlation function
remains well defined. However, it no longer corresponds to a projection matrix,
which is the requested input to the process. Prior to the addition of interactions
all the eigenvalues of the correlation function were either 0 or 1, corresponding to
occupied and unoccupied states. One can think of this as an occupation gap of
exactly $1$. As interactions are gradually increased, and as long as there is no
phase transition, the correlation function is expected to change continuously, and
the occupation gap to remain finite. Provided that, it is possible to extract a
truncated projector out of the correlation function in a way which is independent of
the truncation process, as will be proven below.

In general, locality of topological order implies a bulk-edge correspondence theorem, provided that the topological order is well defined for insulating phases. Assume that two samples which belong to two different classes are attached. If the entire system remains gapped, it belongs to a single topological
class. However, according to the locality of the order, the system seems to belong to two classes simultaneously. This contradiction implies a closure of the gap, which can only take place at the interface.  In the other way, if the gap can remain open between the two attached samples, the classifying order must be nonlocal. For example, weak topological insulators may have a gapped surface (in a stacking direction) \cite{FUKANEMELE}, which means that the weak order is nonlocal, in agreement to what has been stated above.

\section{Numerical implementations} 
\label{Sec:Numerics}

\begin{figure}[t]
\begin{center}
\vspace{0cm}
\includegraphics[width=8cm,height=8.2cm,angle=0]{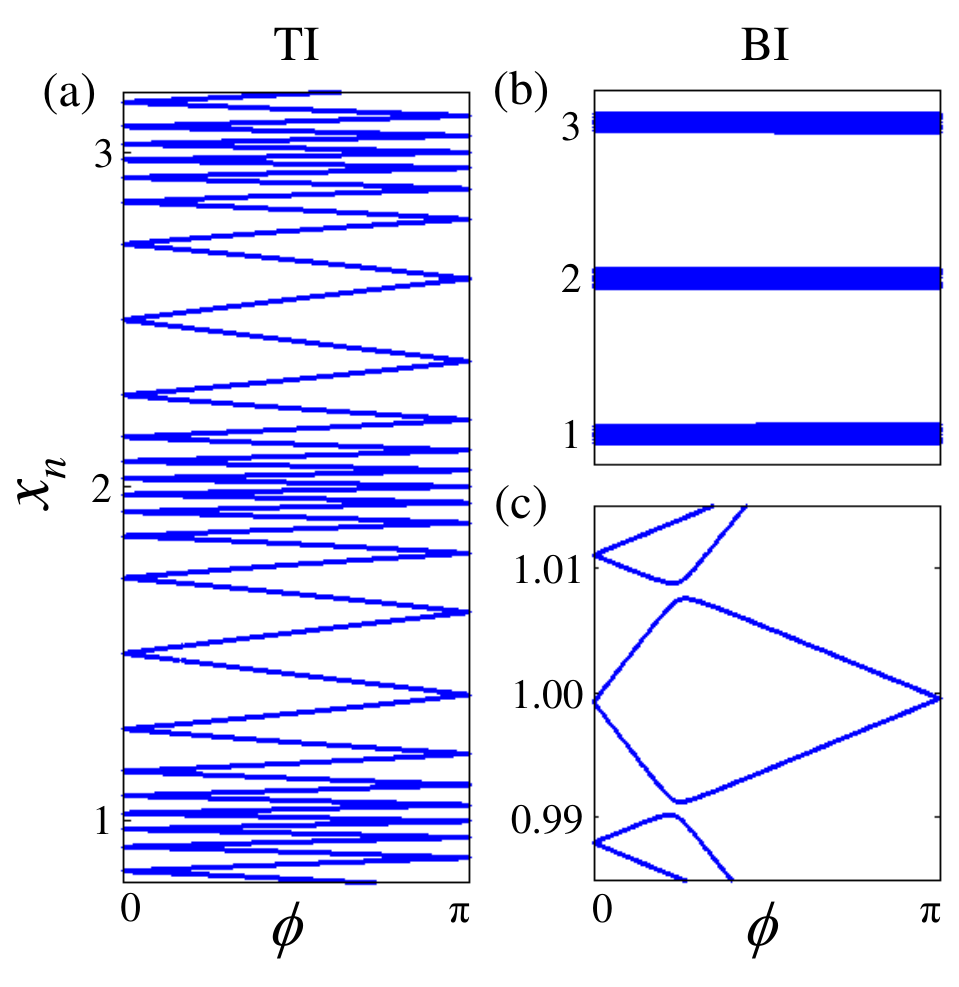}
\vspace{0cm} \caption{ \label{Fig:Switching} %
The centers of the 1D Wannier functions $x_n$ as a function of the artificial flux
$\phi$ of the 2D Kane and Mele model, as extracted from the local two-point
correlation function at $\phi = 0$, of both a topological (left) and a trivial
(right) insulators. (a) In the TI the centers switch pairs between $\phi = 0$ and
$\phi = \pi$. (b) In the BI the centers are essentially fixed. (c) Zooming-in shows
that they actually move, but without switching pairs. }
\end{center}
\end{figure}

The above approach can be easily adapted to a computer algorithm. For a given
two-point correlation function $P$, one can diagonalize the matrix $PXP|_{\phi}$ for
various fluxes, and examine the resulting eigenvalue spectrum. However, due to
finite size effect, some of the eigenstates of $PXP$ would be localized near the
edges of the region $\A$. These eigenstates are edge dependent and,
therefore, do not reflect the physical behavior of the bulk. Fortunately, due to the
exponential localization of the eigenstates, one can easily distinguish these state
from the bulk eigenstates and discard them, as proven in the Appendix.

We have carried out this process on the 2D Kane and Mele model of the TI
\cite{KANEMELE} in the presence of disorder. The Hamiltonian of this model is
parameterized by nearest neighbor hopping $t$, staggered on-site potential
$\lambda_v$, $S_z$ conserving spin-orbit interaction $\lambda_{SO}$, and Rashba
spin-orbit interaction $\lambda_R$. We took a cylindrical $18 \times 42$ lattice,
with $t = 1, \lambda_{SO} = 0.1, \lambda_R = 0.05$, for the BI $\lambda_v = 0.9$,
while for the TI $\lambda_v = 0.1$. Both Hamiltonians were subject to a uniformly
distributed random potential of magnitude $0.1$. The spectral flows $x_n(\phi)$ at
the middle of the cylinder of both the trivial and topological phases are depicted
in Fig.~\ref{Fig:Switching}. The difference in the pair switching behavior is
clearly visible.

\begin{figure}[t]
\begin{center}
\vspace{0cm}
\includegraphics[width=8cm,height=4cm,angle=0]{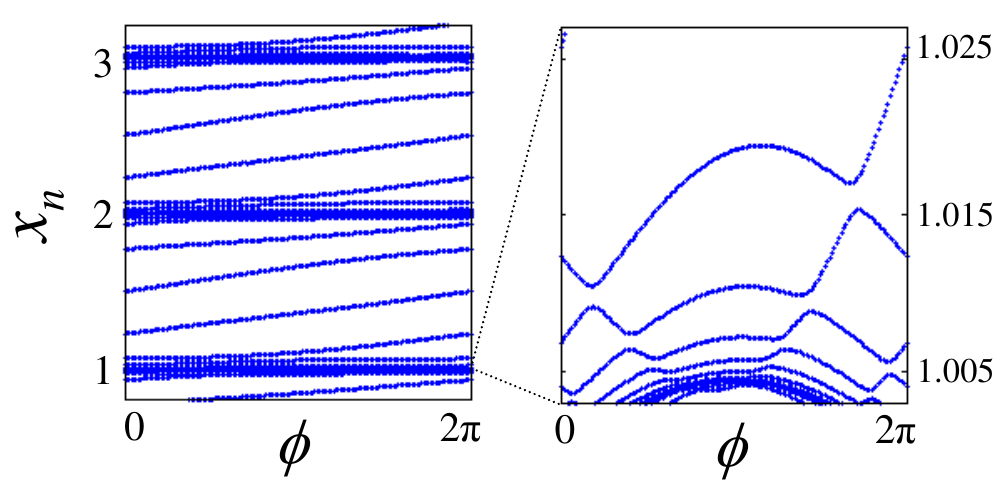}
\vspace{0cm} \caption{ \label{Fig:Z_index} %
The centers $x_n$ versus the artificial flux $\phi$ of a single spin copy of the
quantum spin Hall effect model, as extracted from $P$ at zero flux. Each center
replaces its proceeding center while carried with the flux, $x_n(2\pi) =
x_{n+1}(0)$, in all scales, as expected from $\nu = 1$. }
\end{center}
\end{figure}

Similarly, we performed the procedure on a single spin copy of the quantum spin Hall
effect model \cite{QSHE}, which is equivalent to IQHE. The parameters of this models
are the hopping element of the two bands $B-D$ and $B+D$, the interband hopping $A$,
and the energy gap $M$. Figure \ref{Fig:Z_index} depicts the movement of the 1D
Wannier functions at the middle of a cylindrical $30 \times 40$ lattice, with $B =
1$, $D = 0.2$, $A = 0.4$, and $M = 0.1$, which yield $\nu = 1$. A uniform disorder of
magnitude $0.3M$ is also present. It is evident that $x_n(2\pi) = x_{n+1}(0)$, which
means that this insulator belongs to class $1$ of the \Y1 index, as expected.

A comment is in order that different $x_n(\phi)$'s may appear to cross at accidental
values between $\phi = 0$ and $\phi = \pi$; for example, see
Fig.~\ref{Fig:Switching}(c). In such cases one should relate the upper state before
the crossing to the upper state after the crossing and the same for the lower. This
process is equivalent to opening a gap at the crossing point by some small local
perturbation. Moreover, due to Winger noncrossing theorem \cite{WIGNER}, this gap
is probably there, only it is not visible within the numerical accuracy.

\section{Proofs of analytical properties} 
\label{Sec:Proofs}

Now we turn to prove the three properties that have been stated above. The band
insulator lattice Hamiltonian $H$ is characterized by a spectrum with a finite gap
$\Delta$ around some value $\mu$, and a maximal energy in absolute value $D$ above
$\mu$. We also assume that $H$ does not couple two sites which are more than $l_h$
sites apart. Since we are interested in bulk properties, we assume that the periodic
dimensions of the lattice are of finite size $L$, while the open coordinate is infinite
(finite size systems are discussed in the Appendix).

The starting point of the proofs is to develop a representation of the projector $P$
as a finite polynomial in $H$. The projector $P$ can also be expressed as
\begin{align} \label{Eq:PofH}
P = \lim_{\epsilon \rightarrow 0^+} \half \left( 1 - \erf\left( \frac{H
-\mu}{\epsilon\Delta} \right) \right),
\end{align}
where $\erf(x)$ denotes the Taylor series of the error function. The equivalence of
this expression to the definition can be easily verified in the eigenbasis of $H$.
Accordingly, we define the approximate projector $\Pe$ to be the same as $P$ but
with finite $\epsilon \ll 1$. If the error of the approximation is measured by the
Euclidean norm, we can bound it by $\| P - \Pe \| = ( 1 - \erf(\epsilon^{-1}) )/2 <
e^{-1/\epsilon^2}$.

Since the $\erf(x)$ is an entire function, $\Pe$ can be approximated by taking only
the $N$ first terms of the series
\begin{align} \label{Eq:PNe}
\PNe = \half \left( 1 - \frac{2}{\sqrt{\pi}}\sum_{n=0}^{N} \frac{(-1)^n}{n!(2n+1)}
\left( \frac{H-\mu}{\epsilon\Delta} \right)^{2n+1} \right).
\end{align}
According to Taylor's theorem there exists $x_0 \in (0,(D/\Delta)\epsilon^{-1})$
such that $\| \Pe - \PNe \| = x_0^{\phantom{e}2N+3} / [\sqrt{\pi}(N+1)!(2N+3)]$. By
using Stirling's approximation, and choosing $N_\epsilon = e^2 (D/\Delta)^2
\epsilon^{-2} \gg \epsilon^{-2}$, we can bound
\begin{align} \label{Eq:PeminPNe2}
\| \Pe - \PNe \| &< \frac{D}{\sqrt{N}\Delta\epsilon} \left( \frac{ e D^2 } { N
\Delta^2 \epsilon^2 } \right)^{N+1} = e^{-N_\epsilon}.
\end{align}

Consequently, we can conclude that $P_{N_\epsilon,\epsilon}$ is an excellent
approximation of $P$, with an error of $\| P - P_{N_\epsilon,\epsilon} \| <
e^{-1/\epsilon^2} \equiv e^{-Q}$, with $Q$ as the measure of the accuracy.
Therefore, in order to approximate $P$ with accuracy $Q$, it is sufficient to take
$N_Q = e^2(D/\Delta)^2Q$ terms.

{\bf Property I:} $P_{ij}$ decays exponentially with the distance between $i$ and
$j$ within some correlation length $l_p$, and has only an exponentially small
dependence on the Hamiltonian outside a local region around $i$ and $j$.

\emph{Proof:} According to the assumptions, $[H^{2N+1}]_{ij}$ vanishes for two sites
$i,j$ which are separated a distance $r_{ij} > (2N+1)l_h \approx 2Nl_h$.
Accordingly, $[\PNe]_{ij}$ also vanishes, while $P_{ij}$ may be of order
$e^{-Q(N)}$. Therefore, for a given $r_{ij}> 2l_h$, we choose $N = [r_{ij}/2l_h]$,
which keeps $[\PNe]_{ij}$ as zero, while giving a maximal $Q(N)$. Following this
choice one finds that
\begin{align} \label{Eq:Pij}
& |P_{ij}| < e^{-r_{ij}/l_p}, \\
& l_p = 2e^2 (D/\Delta)^2 l_h.
\end{align}

Moreover, $P_{ij}$ for two sites within a ceratin region $\A$ does not
depend on the matrix elements $H_{kl}$ of two sites within region $\mathcal{B}$ as
long as the distance between $\A$ and $\mathcal{B}$ is much larger than
$l_p$. This can be shown by choosing $N = r_{\mathcal{AB}}/2l_h$, where
$r_{\mathcal{AB}}$ is the minimal distance between $\A$ and $\mathcal{B}$.
Now $[\PNe]_{ij}$ does not depend on $H_{kl}$, and $P_{ij}$ may depend on it on the
order of $e^{-r_{\mathcal{AB}}/l_p}$ \cite{HASTINGS}.

This way of creating an exponentially local projector from the Hamiltonian can be
performed on any gapped matrix with a finite band width. In particular the two-point
correlation function of a weakly interacting system can be deformed into a
projector, as required for the index determining procedure, as long as the
occupation gap remains finite.

{\bf Property II:} Magnetic flux which threads a cylinder or a torus affects
$P_{ij}$ by a simple geometric phase factor.

\emph{Proof:} Insertion of the magnetic flux $\Phi$ affects $H$ in the uniform
gauge by
\begin{align} \label{Eq:HofPhi}
H_{kl}(\Phi) = H_{kl} e^{ i2\pi y_{kl}\Phi/L \Phi_0},
\end{align}
where $y_{kl} \in (-L/2, L/2]$ is the azimuthal distance from site $k$
to site and $l$, and $\Phi_0$ is the flux quanta. Consider $H_{kl}$ within a
region $\mathcal{C}$ that cover less than a half of the circumference of the cylinder
or the torus. One can adopt a convention for the coordinate $y \in (0,L)$ that
avoids the branch cut inside $\mathcal{C}$ and write $y_{kl} = y_k - y_l$.
Within $\mathcal{C}$ Eq.~(\ref{Eq:HofPhi}) can now be written as a unitary transformation
\begin{align} \label{Eq:HofU}
& H(\Phi)|_\mathcal{C} = U(\Phi) H U\da(\Phi)|_\mathcal{C}, \\
& U_{kl}(\Phi) = \delta_{kl} e^{ i2\pi y_k\Phi/L \Phi_0}, \qquad k,l \in
\mathcal{C}.
\end{align}

Given $P_{ij}$, we can approximate it by $N = L/4l_h$ terms. Since $[\PNe]_{ij}$
does not vanish only for $r_{ij} \sim l_p \ll L$, it depends on $H_{jk}$ only for
$j,k$ within a region of size $L/2$, which covers not more than a half of the
circumference. Hence we may substitute Eq.~(\ref{Eq:HofU}) in Eq.~(\ref{Eq:PNe})
and obtain Eq.~(\ref{Eq:PofPhi}) by identifying $\phi = \Phi/\Phi_0$ and
$\theta_{ij} = 2\pi y_{ij}/L$.

{\bf Property III:} The 1D Wannier functions, defined as eigenstates of the operator
$PXP$, are exponentially localized in the $\hat{x}$ direction around the
corresponding eigenvalues.

\emph{Proof:} Consider the action of the projector $P$ on some normalized state
$|a\rangle$ which is localized around $x_a$ within a width $l_a$. Since $P_{ij}$ is
local with width $l_p$, $P|a\rangle$ is still localized at the vicinity of $x_a$,
but it might be as wide as $l_a + l_p$. Note that $P|a\rangle$ is not necessarily
normalized to $1$ but may have a smaller norm. Since the position operator $X$ is
diagonal in the position basis, $(X - x_a)|a\rangle$ is localized almost in the same
manner as $|a\rangle$, but its norm may increases up to $2l_a$.

Following this, if we begin with a particle at site $i$, and apply $P(X - x_i)P$ on
it, then $\| P(X - x_i)P|i\rangle \| < 2l_p$. Applying it $M$ times yields $\| (P(X
- x_i)P)^M|i\rangle \| < (2l_p)^M M! \approx \sqrt{2\pi M} \left( 2l_pM/e
\right)^M$.

Now let $|w_n\rangle$ be an eigenstate of $PXP$ with eigenvalue $x_n \neq 0$, and
consider $|i\rangle$ with $|x_n - x_i|> 2l_p$. It can be shown that $P|w_n\rangle =
|w_n\rangle$, yielding $\langle i| w_n \rangle = (x_n-x_i)^{-M} \langle
i|(P(X-x_i)P)^M|w_n\rangle$. Following the inequality
$| \langle a| b \rangle | \le \| |a\rangle \| \cdot \| |b\rangle \|$, we have
$| \langle i| w_n \rangle | < \sqrt{2\pi M} \left[ 2l_pM /(e(x_n - x_i)) \right]^M$.
For given $x_n$ and $x_i$ we choose $M = [ |x_i - x_n|/2l_p ]$, and get the exponential
localization of the 1D Wannier function \cite{NENCIU}
\begin{align} \label{Eq:iWn3}
| \langle i| w_n \rangle | < \sqrt{\pi|x_i - x_n|/l_p} \, e^{-|x_i - x_n|/2l_p}.
\end{align}

\section{Conclusion} 
\label{Sec:Conclusion}

To conclude, in this work we proved that the \Y1 and \Z2 topological indices can be extracted from the ground-state equal-time two-point correlation function at zero flux, given on any section larger than some correlation length. This implies that the order in such topological phases can be thought of as a local ground-state property. In the case of 3D TI it was suggested that the strong index is indeed local in the above sense, while the weak indices carries some global information.

Heuristically, one can reach a similar conclusion using entanglement spectrum \cite{HALDANE}.
Indeed, $P_{ij}$ on a finite region is related to the reduced density matrix.
A gapless spectrum of $P_{ij}$ is therefore an indication of nontrivial topology
\cite{ASHVIN2,BERNEVIG}. Nevertheless, given a finite region one cannot differentiate a
gapped spectrum from a gapless one without inserting fluxes.

It will be interesting to establish this approach in interacting systems
and to apply it on fractional quantum Hall states. Preliminary numerical results
indicate that the electron two-point correlation function of the ground state at filling factor $1/3$
is proportional to the one of filling factor $1$. Hence, it may be possible
to extract the topological order of fractional states in a similar way.

We hope that our viewpoint will encourage the efforts for finding manifestations
of the \Z2 index through local bulk properties.

\section*{ACKNOWLEDGEMENTS} 

We thank Ady Stern and Ehud Altman for useful discussions. Z.R. thanks
A.~Soroker, and Y.E.K. thanks M.~Kraus. Z.R. Acknowledges ISF Grant No.~700822030182, and Y.E.K. thanks the U.S.-Israel Binational Science Foundation and the Minerva foundation for financial support.

\appendix
\section{Edge effects} 
\label{App:Edges}

In the main text we avoided edge effects by considering geometries with periodic
dimensions and an infinite open coordinate. In practice, a finite open coordinate must
be used, causing numerical artifacts that should be filtered out in order to reveal the bulk
behavior. This can be done in a controlled manner provided that the 1D Wannier functions
of the bulk and the edge are distinguishable. The first part of this Appendix
establishes this distinction, and the second part presents the actual numerical
procedure.

The starting point of the proofs in Sec.~\ref{Sec:Proofs} was expanding the spectral projector $P$ as a
series in powers of the Hamiltonian $H$, with the spectral gap $\Delta$ as a control
parameter. An edge may give rise to gapless edge states and so our first
task is to establish the expansion in the presence of such states.

We assume that $H$ is characterized as before, only now the
spectrum of $H$ includes both bulk states with energy greater than the gap, and
edge states with subgap energy, which are exponentially localized along the edge.
The projector $P$ can than be divided into bulk and edge parts
\begin{align} \label{Eq:P_bulk_edge}
P & = P^{bulk} + P^{edge} \\
  & = \sum_{E_n < \mu - \Delta} | n \rangle \langle n |
    + \sum_{\mu - \Delta < E_n < \mu} | n \rangle \langle n |, \nonumber
\end{align}
where $|n\rangle$ denotes an eigenstate with eigenvalue $E_n$, and $\mu$ is the
chemical potential, as before. Since $P^{edge}$ is composed only of the edge states,
it decays exponentially into the bulk. Therefore $[P^{edge}]_{ij}$ is exponentially
small, if either $i$ or $j$ (or both) reside within the bulk.

In a similar manner to what we have done above, we introduce an error function approximation
to the projector
\begin{align} \label{Eq:Pe_bulk_edge}
\Pe & = P^{bulk}_{\epsilon} + P^{edge}_{\epsilon} \\
    & = \sum_{E_n < \mu - \Delta} \half \left( 1 - \erf\left( \frac{E_n
-\mu}{\epsilon\Delta} \right) \right) | n \rangle \langle n | \nonumber \\
    & + \sum_{\mu - \Delta < E_n < \mu} \half \left( 1 - \erf\left( \frac{E_n
-\mu}{\epsilon\Delta} \right) \right) | n \rangle \langle n |.
\end{align}
We can see that $\| P^{bulk} - P^{bulk}_{\epsilon} \| < e^{-1/\epsilon^2}$ as
before, while $P^{edge}_{\epsilon}$ is a poor approximation to $P^{edge}$, since the
summation coefficients spread form 0 to 1. Indeed $P^{edge}_{\epsilon}$ is a poor
approximation of $P^{edge}$ at the edge. However, within the bulk both
$[P^{edge}_{\epsilon}]_{ij}$ and $[P^{edge}]_{ij}$ are exponentially small, and
therefore also the error $| [P^{edge}_{\epsilon} - P^{edge}]_{ij} |$.

This implies that for bulk-bulk or bulk-edge correlation, $P$ may still be
approximated by $P_{\epsilon}$ within an exponential accuracy. $P_{\epsilon}$ can be
Taylor expanded up to some finite order $N$, and by choosing optimal values for
$\epsilon$ and $N$ we can bound the error. Since this part of the proof is
unaffected by the edges, we refer the reader back to Sec.~\ref{Sec:Proofs}.
We turn to discuss the effect of edges on properties I--III.

\vspace{5mm}

Property I (the exponential locality of $P$) relies on the serial
expansion, and therefore is valid only within the bulk. Nevertheless, since the edge
states decay into the bulk, $P$ decays exponentially also at the vicinity of the
edge but only perpendicularly to the edge.

Property II (the artificial flux insertion) is valid as long as $P$ decays along the periodic
coordinates, which does not necessarily happen in the presence of edge states. Thus
the artificial flux insertion approximation is valid only far from the edge.

Property III states that the 1D Wannier functions are exponentially localized around
their eigenvalues along the open coordinate $X$. As long as the 1D Wannier functions
are produced by diagonalizing $PXP$, and $P$ is a true projection operator over the entire
physical system, the proof remains unchanged in the presence of edges. Nevertheless, we
do not wish to limit ourselves to cases in which the entire $P$ matrix is known, since
this is not a local quantity. Our algorithm uses $P$ which is given on some local area
with a geometry of a cylinder or a Corbino disk. But $P$ which is truncated to some local
region is generally not a projection matrix.

The projection property $P^2 = P$ was used in the original proof only once,
when it was stated that it can be shown that $P |w_n \rangle = | w_n \rangle$, where
$|w_n\rangle$ is an eigenfunction of $PXP$ with eigenvalue $x_n \neq 0$. This property
was required in order to establish the equality
\begin{align} \label{Eq:PIII}
\langle i | w_n \rangle = (x_n-x_i)^{-M} \langle i | (P(X-x_i)P)^M |w_n\rangle,
\end{align}
where $|i\rangle$ is a state localized at site $i$ . If the truncated $P$ is no longer
a projector, then $[PXP, P] \neq 0$, and $|w_n \rangle$ may not be an
eigenstate of $P$. But since we are interested in bulk properties, it  suffices to
prove that Eq.~(\ref{Eq:PIII}) is valid far from the edges of $P$ (which may
differ from the physical edges due to the truncation).

It is useful to
divide the the system into three groups: $\mathcal{S}$ will denote the entire
system, $\A \subset \mathcal{S}$ the truncation area, and $\B \subset \A$ the inner
region of $\A$, which will be defined by the set of all sites which are far form the
edges of $\A$, on the scale of the the correlation length $l_p$. Recall that $P$ is
a piece of the true projector of the entire system, which we denote by $\tilde{P}$.
Following the locality of $P$, we find that for $i$ or $j$ in $\B$
\begin{align} \label{Eq:AlmostP}
[P^2]_{ij} &= \sum_{k \in \A} P_{ik}P_{kj} = \sum_{k \in \A} \tilde{P}_{ik}
\tilde{P}_{kj} \nonumber  \\
&\approx \sum_{k \in \mathcal{S}} \tilde{P}_{ik}\tilde{P}_{kj} = \tilde{P}_{ij} =
P_{ij},
\end{align}
up to corrections that are exponentially small in the distance between the site in
$\B$ and the edge of $\A$, divided by $l_p$. We can see that $P$ is still a projector
up to boundary effects.

Since $P$ is approximately a projector within $\B$,
\begin{align}
\langle i| P |w_n \rangle & = (1/x_n) \langle i| P P X P | w_n \rangle  \\
& \approx (1/x_n) \langle i | P X P |w_n\rangle = \langle i|w_n \rangle \quad
\forall i \in \B, \nonumber
\end{align}
for $x_n \neq 0$. This means that as far as region $\B$ is considered, $|w_n\rangle$
is indeed an eigenstate of $P$. Seemingly we recovered Eq.~(\ref{Eq:PIII}) for any
site $i$ within $\B$. However, $(P(X-x_i)P)^M |i \rangle$ may be as wide as $M
\cdot 2l_p$, which restricts the validity of Eq.~(\ref{Eq:PIII}) to $M <
d(i,\B)/2l_p$, where $d(i,\B)$ is the distance between site $i$ and the edge of
$\B$.

The exponential localization of $|w_n \rangle$ around $x_n$ was achieved by choosing
$M = [ |x_i - x_n|/2l_p ]$. The restriction on $M$ is then translated to a
restriction on the exponential localization to be valid only for $|x_i - x_n| <
d(i,\B)$, although $x_i$ and $x_n$ are both in $\B$.

So far we have shown that for $x_n$ in $\B$, its eigenfunction $|w_n\rangle$ decays
exponentially with the distance. Note that this does not exclude the possibilities
that $|w_n\rangle$ resides at the edge of $\A$, or both at the edge and around $x_n$.
The last scenario is, however, highly nongeneric. Assume that some $|w_n\rangle$ is
indeed doubly localized in that fashion. Up to exponential accuracy we may split
it into sum of two functions $|w_n \rangle = |w_n,\B \rangle +
|w_n,\A \rangle$, where $|w_n,\B \rangle$ is localized around $x_n$ and $|w_n,\A \rangle$
is localized around the edge. Since $PXP$ is local, it does not couple these two
functions, and $|\tilde{w}_n \rangle = |w_n,B\rangle -  |w_n,A\rangle$ is also an
eigenfunction, with an eigenvalue that is exponentially close
to $x_n$. Such an almost degeneracy is of course nongeneric.

\begin{figure}[ht!]
\begin{center}
\vspace{0cm}
\includegraphics[width=8cm,angle=0]{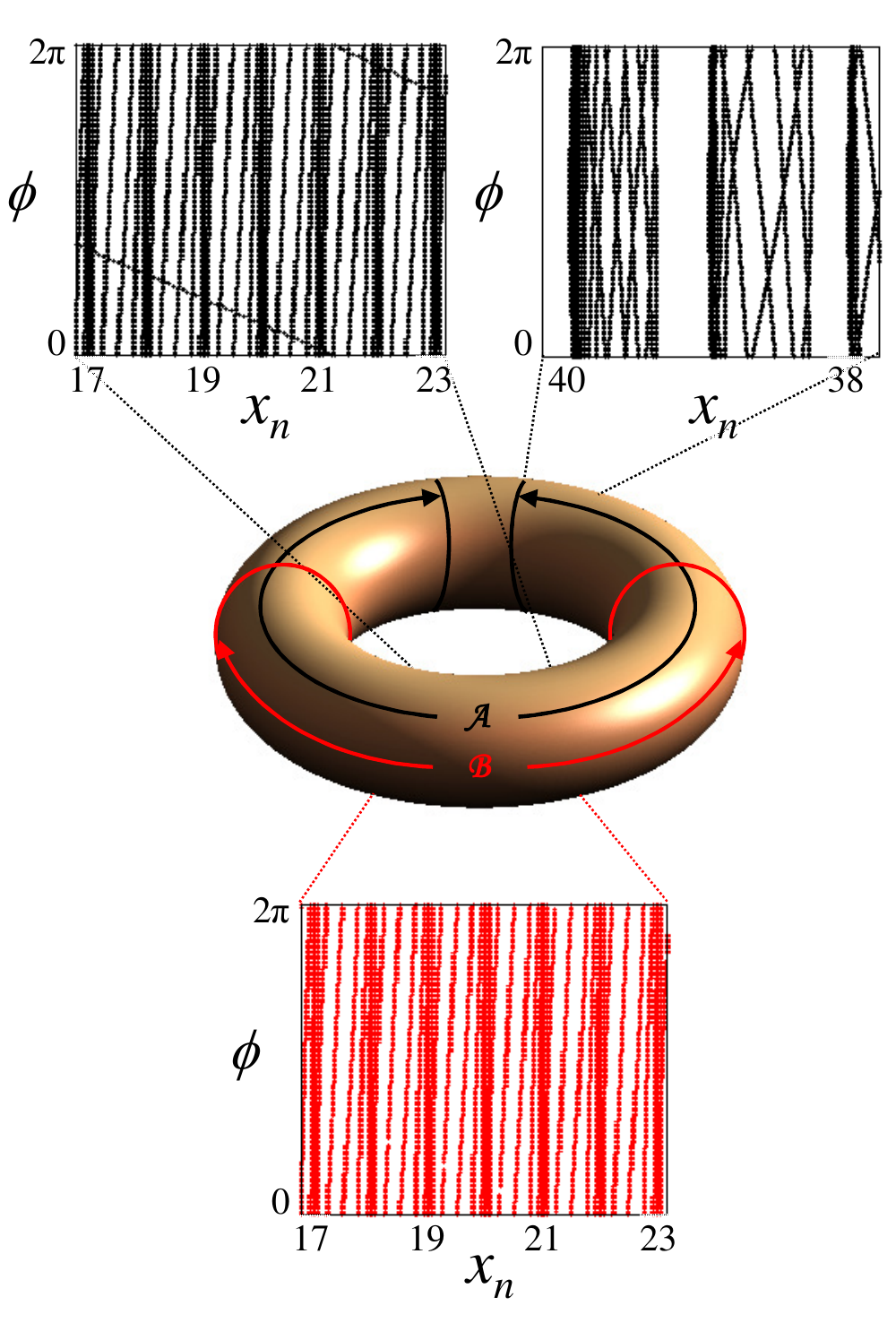}
\vspace{0cm} \caption{ \label{Fig:Edge_effect} %
Demonstration of bulk-edge separation in the single spin copy of the quantum spin
Hall effect. The projector $P$ was created by truncating a $30 \times 34$ cylinder,
denoted by $\A$, from the $30 \times 40$ torus. The full spectrum of $PXP|_{\phi}$
(top, black) is trivial both in the bulk and at the edge.
Omitting states that are localized at the edges gives a pure bulk spectrum at
region $\B$ (bottom, red), which is indeed nontrivial. }
\end{center}
\end{figure}

To conclude, we have seen that edges do not affect the bulk properties of $P$.
Moreover, the 1D Wannier function with eigenvalue in the bulk are localized either
around their eigenvalue or at the edge. Therefore, given the correlation function $P$
on an arbitrary geometry, one can isolate/truncate a cylinder or a Corbino disk from it and
distinguish the 1D Wannier functions that reside within the bulk . Due to their
localization properties, the 1D Wannier functions are unaffected by the edges or the
truncation process, and the motion of their eigenvalues as a function of the flux
is therefore purely a bulk property.

\vspace{5mm}

We now demonstrate how these analytical statements are exploited in the computer
algorithm. For that purpose we fully diagonalized the Hamiltonian of the single spin
copy of the quantum spin Hall effect, for a periodic $30 \times 40$ lattice, with
the parameters that are given in Sec.~\ref{Sec:Numerics}. The projector $P$ was than produced
by summing all projectors on the states with negative energy and was characterized
by a correlation length of approximately 2. Region $\A$ was chosen be a $30 \times 34$
cylinder out of the full torus, and all the elements $P_{ij}$ with $i$ and $j$ out
of $\A$ were omitted.

The next step was to construct the spectrum $\{x_n(\phi)\}$ by diagonalizing $PXP|_{\phi}$,
where $X$ in region $\A$ ranged from 4 to 37. The full spectrum is trivial, as depicted at
the top of Fig.~\ref{Fig:Edge_effect}. At the edges of $\A$ edge states
hybridize with the outmost bulk states, and gaps are opened. At the center of $\A$
most eigenvalues belong to bulk states, while the branch of eigenvalues that
crosses the spectrum from side to side belongs to an edge state.

In order to remain with bulk effects only, we used the localization property of the
1D Wannier functions, which guarantees that the bulk wave function are localized within
the middle of region $\A$, denoted above by $\B$. We chose $\B$ to be the central
cylinder of 20 sites, and discarded all the 1D Wannier function that more than
5\% of their weight is outside $\B$. In this way we assured that all the states in $\B$
are purely bulk states, and the nontrivial nature of the spectrum became apparent,
as seen in the bottom of Fig.~\ref{Fig:Edge_effect}.


\begin{thebibliography}{99}
\bibitem{Chern}
Consequently, using the above equation for two fluxes in the expression for the Chern number,
$\nu = \frac{1}{2\pi i}\int d\phi_x d\phi_y \textrm{Tr}( P,[\partial_{\phi_x}P, \partial_{\phi_y}P] )$,
reveals that the integration over the flux is superfluous.
\end{thebibliography}
\end{document}